\documentclass[12pt]{article}

\usepackage{epsfig}

\begin{document}

\title{Madelung Representation and Exactly Solvable Schr\"{o}dinger-Burgers Equations with Variable Parameters}

\maketitle

\begin{center}
{\bf \c{S}irin A. B\"{u}y\"{u}ka\c{s}{\i}k, Oktay K. Pashaev}\\
Dept. of Mathematics, Izmir Institute of Technology, \\
35430 Urla, Izmir, Turkey\\
sirinatilgan@iyte.edu.tr, oktaypashaev@iyte.edu.tr
\end{center}

\begin{abstract}
                We construct a Madelung fluid model
                 with specific time variable parameters as
                 dissipative quantum fluid and linearize it in terms
                 of Schr\"{o}dinger equation with time dependent parameters.
                 It allows us to  find exact solutions of the nonlinear Madelung
                 system in terms of solutions of the Schr\"odinger equation and
                 the corresponding classical linear ODE with variable frequency and damping.
                 For the complex velocity field  the Madelung system  takes the form of a nonlinear
                  complex Schr\"{o}dinger-Burgers equation, for which we obtain exact solutions
                  using complex Cole-Hopf transformation. In particular, we discuss and give exact
                   results for nonlinear Madelung systems related with Caldirola-Kanai type dissipative
                    harmonic oscillator.
\end{abstract}

\section{Introduction}
In the recent
years the Madelung fluid description of quantum mechanics  has been applied to some fields where the quantum formalism is
a useful tool for describing the evolution of classical
(quantum-like) systems and studying the dispersionless or semiclassical limit of  nonlinear partial differential
equations of  Schr\"odinger type, \cite{Zakharov}.
The Madelung fluid representation, proposed first by
\cite{Madelung}, being a complex quantity, represents a solution of the Schr\"{o}dinger
equation,  in terms of
modulus and phase. Substituted to the Schr\"{o}dinger equation
it allows to obtain a pair of nonlinear hydrodynamic type equations.
Thus, the Madelung fluid equations are nonlinear system of PDEs,
 while the Schr\"{o}dinger equation is the linear one. Then, the Madelung transform is a complex linearization
  transform, similar to the Cole-Hopf transformation, linearizing the nonlinear Burgers equation in terms of the linear heat equation, see \cite{Cole} and \cite{Hopf}.
  Nonlinear models admitting such type of direct linearization are called by F. Calogero as C-integrable models.

  In this work, we construct a Madelung fluid model with time variable parameters as dissipative quantum fluid and linearize it in terms of Schr\"{o}dinger equation with time dependent parameters. It allows us to find exact solution of the nonlinear Madelung system in terms of solutions to the Schr\"odinger equation and the corresponding classical linear ODE with variable frequency and damping. Moreover, the Madelung system written for the complex velocity field takes the form of a nonlinear  complex Schr\"{o}dinger-Burgers equation, which exact solutions we obtain
 using complex Cole-Hopf transformation. As known, in the usual Cole-Hopf transformation zeros of the linear heat equation lead to poles in the corresponding Burgers equation. Similarly, in our case, by the complex Cole-Hopf
transformation  zeros of the Schr\"{o}dinger equation transform to pole singularities in the complex Schr\"{o}dinger-Burgers equation. Thus, using exact solutions of the linear problem, one can find also the dynamics of the poles in the corresponding nonlinear problem.
As an exactly solvable model, we describe a  dissipative nonlinear complex Schr\"{o}dinger-Burgers equation of Caldirola-Kanai type, \cite{Caldirola}, \cite{Kanai}. Exact solutions of the nonlinear models are found and the motion of zeros and poles is  discussed explicitly. Some illustrative plots are constructed.

\section{  The Schr\"{o}dinger Equation and its Madelung
Representation}
\subsection{Solution of the Schr\"{o}dinger Equation }

Consider the one-dimensional Schr\"{o}dinger equation for harmonic oscillator with time-dependent parameters
\begin{equation}\label{evo1}
i\hbar\frac{\partial \Psi}{\partial
t}=-\frac{\hbar^{2}}{2\mu(t)}\frac{\partial^{2} \Psi}{\partial
q^2}+\frac{\mu(t)\omega^{2}(t)}{2}q^{2}\Psi,
\end{equation}
and initial condition
\begin{equation}\label{ic1}
\Psi(q,t_{0})=\psi(q),\,\,\,\,\,\,-\infty<q<\infty.
\end{equation}
Using the Evolution operator method, \cite{Wei-Norman}, it was proved that,
see \cite{S.O.E}, if $x(t)$ is the solution of the classical
equation of motion
\begin{equation}\label{clas-eq}
\ddot{x}+\frac{\dot{\mu}(t)}{\mu(t)}\dot{x}+\omega^{2}(t)x=0,\,\,\,x(t_{0})=x_{0}\neq
0,\,\,\,\dot{x}(t_{0})=0,
\end{equation}
then the solution of the IVP
(\ref{evo1})-(\ref{ic1}) is found as
  $ \Psi(q,t)=\widehat{U}(t,t_{0})\psi(q),$
   where the evolution operator is
\begin{eqnarray*}\label{ut}
\widehat{U} =
\exp
\left(\frac{i}{2}fq^{2}\right)\,\exp\left(
h(q\frac{\partial}{\partial q}+\frac{1}{2})\right)\,
\exp\left(-\frac{i}{2}g\frac{\partial^{2}}{\partial q^2}\right)
\end{eqnarray*}
and the auxiliary functions are
$$
f(t)=\frac{\mu(t)}{\hbar}\frac{\dot{x}(t)}{x(t)};\,\,\,g(t)=-\hbar
x^{2}(t_{0})\int^{t}\frac{d\xi}{\mu(\xi)x^{2}(\xi)},\,\,\,\,g(t_{0})=0;\,\,\,
h(t)=\ln\frac{|x(t_{0})|}{|x(t)|}.$$ In particular, if the initial
function is  the normalized eigenstate corresponding to eignenvalue
$E_{k}=\hbar^{2}\Omega_{0}(k+1/2)$ of the Hamiltonian for the
standard harmonic oscillator, that is
\begin{equation}\label{Ho-ef}
\varphi_{k}(q)=N_{k}
e^{-\frac{\Omega_{0}}{2}q^{2}}H_{k}(\sqrt{\Omega_{0}}q),\,\,\,\,\,k=0,1,2,...,
\end{equation}
then, the time-evolved state for the Schr\"{o}dinger equation (\ref{evo1}) is
\begin{eqnarray}\label{utvar1}
\Psi_{k}(q,t)&=& \hat{U}(t,t_{0})\varphi_{k}(q)\nonumber\\  &=& N_{k}\sqrt{R(t)}\times\exp\left(i\left(k+\frac{1}{2}\right)\arctan(\Omega_{0}g(t))\right) \nonumber \\
& & \times
\exp\left(i\left(
\frac{\mu(t)\dot{x}(t)} {2\hbar x(t)}
-\frac{\Omega_{0}^{2}}{2}g(t) R^{2}(t)\right)q^{2}\right)\nonumber \\
& & \times \exp\left(-\frac{\Omega_{0}}{2}R^{2}(t)q^{2}\right)\times
H_{k}\left(\sqrt{\Omega_{0}}R(t)q\right).
\end{eqnarray}
where
 \begin{equation}\label{Rt} R(t)=
\left(\frac{x_{0}^{2}}{x^{2}(t)+(\Omega_{0}
x(t)g(t))^{2}}\right)^{\frac{1}{2}}.\end{equation}
  The corresponding probability density  is then
  \begin{eqnarray}\label{density}
\rho_{k}(q,t)=  \frac{1}{2^{k}k!\sqrt{\pi}}\times
\sqrt{\Omega_{0}}R(t) \times\exp\left(-\left(\sqrt{\Omega_{0}}R(t)
q\right)^{2}\right)\nonumber  \times H
_{k}^{2}\left(\sqrt{\Omega_{0}}R(t)q\right).
\end{eqnarray}

\subsection{ Madelung representation}

 As known,  Madelung representation of the complex-valued wave function
\begin{equation}\label{Mad}
\Psi(q,t)=\sqrt{\rho}\exp\left(\frac{i}{\hbar}S\right)=\exp\left(\frac{1}{2}\ln\rho+\frac{i}{\hbar}S\right),
\end{equation}
where $\rho=\rho(q,t)$ is the probability density and $S=S(q,t)$ is the
action, both being real-valued functions, decomposes the
Schr\"{o}dinger equation (\ref{evo1}) into a system of nonlinear
coupled  partial differential equations,
\begin{eqnarray*}
 \frac{\partial S}{\partial
t}+\frac{1}{2\mu(t)}\left(\frac{\partial S}{\partial
q}\right)^{2}+\frac{\mu(t)\omega^2(t)}{2}q^2=\frac{\hbar^2}{2\mu(t)}\left[\frac{1}{\sqrt{\rho}}\frac{\partial^2
\sqrt{\rho}}{\partial q^2}\right]
\end{eqnarray*}
\begin{eqnarray}\label{hamilton-jacobi}
 \frac{\partial\rho}{\partial
t}+\frac{\partial}{\partial
q}\left[\rho\frac{1}{\mu(t)}\frac{\partial S}{\partial q}\right]=0.
\end{eqnarray}

 The first equation may be viewed as a
generalization of the usual Hamilton-Jacobi equation. The term with
explicit $\hbar$ dependence is the quantum potential
encoding the quantum aspects of
the theory. When $\hbar\rightarrow 0,$ the equation becomes
Hamilton-Jacobi equation for a non-relativistic particle with time dependent mass. The second
equation is a continuity equation expressing the conservation of
probability density.
 Using the relation (\ref{Mad}), one can see
that the system (\ref{hamilton-jacobi}), with general initial
conditions
\begin{equation}\label{ham-ic}
S(q,t_{0})=\tilde{S}(q),\,\,\rho(q,t_{0})=\tilde{\rho}(q),
\end{equation}
 $\tilde{S}(q),\tilde{\rho}(q)$ being real-valued functions, has
formal solution
\begin{equation}\label{s-prob}
S(q,t)=-i\hbar
\ln\left(\frac{\Psi(q,t)}{|\Psi(q,t)|}\right),\,\,\rho(q,t)=|\Psi(q,t)|^{2},
\end{equation}
 where $\Psi(q,t)$ is a solution of
the Schr\"{o}dinger equation (\ref{evo1}) with  initial condition
\begin{equation}\label{icm}
\Psi(q,t_{0})=\sqrt{\tilde{\rho}(q)}\exp(\frac{i}{\hbar}\tilde{S}(q)).
\end{equation}
We remark that, since $\Psi(q,t)$ is complex-valued, in general $S(q,t)$ is multi-valued, i.e.
$S(q,t)=-i\hbar\ln(\Psi/|\Psi|)+2\pi n \hbar,$ $n=0,\pm 1,\pm 2,...,$  but fixing the initial condition
$S(q,t_{0})=\tilde{S}(q)$ leads to a single-valued solution of the IVP.

\subsection{Madelung Hydrodynamic Equations}
Introducing  classical velocity,
$v(q,t)=\displaystyle\frac{1}{\mu(t)}\frac{\partial S}{\partial
q}\,,$ the system (\ref{hamilton-jacobi}) transforms to Madelung fluid equations
\begin{eqnarray}\label{clas-velocity}
\left\{
\begin{array}{ll}
\displaystyle \frac{\partial v}{\partial t}+ \frac{\dot{\mu}(t)}{\mu(t)}v+v\frac{\partial
v}{\partial q}
=-\frac{1}{\mu(t)}\frac{\partial}{\partial
q}\left[\frac{-\hbar^2}{2\mu(t)}\left(\frac{1}{\sqrt{\rho}}\frac{\partial^2
\sqrt{\rho}}{\partial q^2}\right)+\frac{\mu(t)\omega^2(t)}{2}q^2\right],\\
\displaystyle \frac{\partial\rho}{\partial
t}+\frac{\partial}{\partial q}\left[\rho v\right]=0.
\end{array}
\right.
\end{eqnarray}
These equations are similar to the classical hydrodynamic  equations where $\rho(q,t)$ is the density and $v(q,t)$
is the velocity field of the one-dimensional fluid.   The system of
fluid equations (\ref{clas-velocity}) with general initial
conditions
$$v(q,t_{0})=\tilde{v}(q),
\,\,\,\rho(q,t_{0})=\tilde{\rho}(q),$$
 $\tilde{v}(q)$,$\,\tilde{\rho}(q)$ being real-valued functions, has formal solution
\begin{equation}
 v(q,t)=-\frac{i\hbar}{\mu(t)}\frac{\partial}{\partial
q}\ln\left(\frac{\Psi(q,t)}{|\Psi(q,t)|}\right),\,\,\rho(q,t)=|\Psi(q,t)|^{2},
\end{equation}
where $\Psi(q,t)$ is solution of the Schr\"{o}dinger equation
(\ref{evo1}) subject to the initial condition
$$\Psi(q,t_{0})=\sqrt{\tilde{\rho}(q)}\exp\left(\frac{i}{\hbar}\mu(t_{0})\int^{q}\tilde{v}(\xi)d\xi\right).$$

\section{ Potential Schr\"{o}dinger-Burgers
 Equation}

Writing the wave function in the form
$$\Psi(q,t)=\exp\left(\frac{i}{\hbar}\mu(t)F(q,t)\right),$$ where
$F(q,t)$ is a complex potential,
 the IVP for the Schr\"{o}dinger equation (\ref{evo1}) transforms to the following IVP for the nonlinear
potential Schr\"{o}dinger-Burgers equation

\begin{eqnarray}\label{pot-burgers-ic}
\left\{
\begin{array}{ll}
 \displaystyle \frac{\partial F}{\partial t}+\frac{\dot{\mu}(t)}{\mu(t)}F+\frac{1}{2}(\frac{\partial
F}{\partial
q})^{2}+\frac{\omega^{2}(t)}{2}q^{2}=\frac{i\hbar}{2\mu(t)}\frac{\partial^{2}F}{\partial
q^{2}}\,\,,\\
F(q,t_{0})=\widetilde{F}(q),
\end{array}
\right.
\end{eqnarray}
Therefore, the formal solution of this problem is
\begin{equation}\label{pot-tr}
F(q,t)=-\frac{i\hbar}{\mu(t)}(\ln \Psi(q,t))\,\,,
\end{equation}
where $\Psi(q,t)$ is  solution of the  Schr\"{o}dinger
equation (\ref{evo1}), with initial condition
$$
\Psi(q,t_{0})=\exp\left(\frac{i}{\hbar}\mu(t_{0})\widetilde{F}(q)\right).$$
 Note again, that fixing the initial condition we obtain a single-valued solution $F(q,t).$
 Now, using the Madelung
representation (\ref{Mad}) and relation (\ref{pot-tr}), one can write
\begin{equation}\label{Fqt}
F(q,t)=\textsf{F}_{1}+i\textsf{F}_{2}=\frac{1}{\mu(t)}S-\frac{i
\hbar}{2\mu(t)}\ln\rho,
\end{equation}
where  $\textsf{F}_{1}=\textsf{F}_{1}(q,t)$ represents the velocity potential, and
$\textsf{F}_{2}=\textsf{F}_{2}(q,t)$ the stream function of the fluid, (
$\textsf{F}_{1},\textsf{F}_{2}$ being real-valued). Accordingly, the
real and imaginary parts of the potential Schr\"{o}dinger-Burgers
equation (\ref{pot-burgers-ic}) become
\begin{eqnarray}\label{F1F2}
\left\{
\begin{array}{ll}
\displaystyle \frac{\partial \textsf{F}_{1}}{\partial
t}+\frac{\dot{\mu}}{\mu}\textsf{F}_{1}+\frac{1}{2}\left((\frac{\partial
\textsf{F}_{1}}{\partial q})^{2}-(\frac{\partial
\textsf{F}_{2}}{\partial
q})^{2}\right)+\frac{\omega^{2}(t)}{2}q^{2}=-\frac{\hbar}{2\mu}\frac{\partial^{2}
\textsf{F}_{2}}{\partial q^{2}},\\
\displaystyle \frac{\partial \textsf{F}_{2}}{\partial
t}+\frac{\dot{\mu}}{\mu}\textsf{F}_{2}+\frac{\partial
\textsf{F}_{1}}{\partial q}\frac{\partial \textsf{F}_{2}}{\partial
q}=\frac{\hbar}{2\mu}\frac{\partial^{2} \textsf{F}_{1}}{\partial
q^{2}}.
\end{array}
\right.
\end{eqnarray}
 Using the relations (\ref{Fqt}) and
(\ref{s-prob}), one can see that the nonlinear system (\ref{F1F2}) with
general initial conditions
$$\textsf{F}_{1}(q,t_{0})=\widetilde{\textsf{F}}_{1}(q),\,\textsf{F}_{2}(q,t_{0})=\widetilde{\textsf{F}}_{2}(q)$$
and $\widetilde{\textsf{F}}_{1}(q)$, $\widetilde{\textsf{F}}_{2}(q)$
real-valued functions, has  solution of the form
$$\textsf{F}_{1}=-\frac{i\hbar}{\mu(t)}\ln\left(\frac{\Psi}{|\Psi|}\right),
\,\textsf{F}_{2}=-\frac{\hbar}{\mu(t)}\ln(|\Psi|),$$
where $\Psi(q,t)$ is  solution of the Schr\"{o}dinger equation
(\ref{evo1}) with  initial condition
$$\Psi(q,t_{0})=\exp\left(\frac{i}{\hbar}\mu(t_{0}) \widetilde{\textsf{F}}_{1}(q)\right)\times \exp\left(-\frac{1}{\hbar}\mu(t_{0})\widetilde{\textsf{F}}_{2}(q)\right).$$

\section{ Schr\"{o}dinger-Burgers
 Equation }

Representation of the wave function in the form
\begin{equation}
\Psi(q,t)=\exp\left(\frac{i}{h}\mu(t)\int^{q}V(\xi,t)d\xi\right),
\end{equation}
where $V(q,t)$ is a complex velocity, transforms the IVP for the Schr\"{o}dinger
equation (\ref{evo1}) to the following IVP for a nonlinear Schr\"{o}dinger-Burgers
equation with time dependent coefficients
\begin{eqnarray}\label{burgers-ic}
\left\{
\begin{array}{ll}
\displaystyle
 \frac{\partial V}{\partial t}+\frac{\dot{\mu}(t)}{\mu(t)}V+V\frac{\partial V}{\partial
q}+\omega^{2}(t)q=\frac{i\hbar}{2\mu(t)}\frac{\partial^{2}V}{\partial
q^{2}}\,\,,\\
V(q,t_{0})=\widetilde{V}(q)\,\,,
\end{array}
\right.
\end{eqnarray}
 Solution of this IVP is found  by the complex Cole-Hopf transformation
\begin{equation}\label{cole-hopf}
V(q,t)=-\frac{i\hbar}{\mu(t)}\frac{\partial}{\partial q}(\ln
\Psi(q,t)),
\end{equation}
where  $\Psi(q,t)$ is  solution of the  the Schr\"{o}dinger
equation (\ref{evo1}), corresponding to initial condition
$$\Psi(q,t_{0})=\psi(q)=\exp\left(\frac{i}{\hbar}\mu(t_{0})\int^{q}\widetilde{V}(\xi)d\xi\right).$$

Using the Madelung representation (\ref{Mad})
and the complex Cole-Hopf transformation (\ref{cole-hopf}) one can write the
complex velocity function in the form
 \begin{equation}\label{Vqt}
 V(q,t)=v+iu=\frac{1}{\mu(t)}\frac{\partial S}{\partial
 q}-\frac{i\hbar}{2\mu(t)}\frac{\partial}{\partial q}(\ln\rho),
\end{equation}
where $v=v(q,t),$ $\,u=u(q,t)$ are real-valued, $v$ represents the classical velocity,
and $u$ the quantum velocity. This splits the
Schr\"{o}dinger-Burgers equation into real and imaginary parts,
respectively,
\begin{eqnarray}\label{hyd-eq}
\left\{
\begin{array}{ll}
\displaystyle \frac{\partial v}{\partial t}+\frac{\dot{\mu}(t)}{\mu(t)}v+v\frac{\partial
v}{\partial q}+\omega^{2}(t)q= \frac{-\hbar}{2\mu(t)}\frac{\partial^{2} u}{\partial
q^{2}}+u\frac{\partial
u}{\partial q},\\
\displaystyle \frac{\partial u}{\partial
t}+\frac{\dot{\mu}(t)}{\mu(t)}u+u\frac{\partial v}{\partial
q}+v\frac{\partial u}{\partial
q}=\frac{\hbar}{2\mu(t)}\frac{\partial^{2}v}{\partial q^{2}}.
\end{array}
\right.
\end{eqnarray}
The first equation is a hydrodynamic equation for the classical
velocity, and the second one is the transport equation for the
quantum velocity.  Using relations (\ref{Vqt}) and
(\ref{s-prob}), we find that the system of nonlinear coupled
equations (\ref{hyd-eq}) with general initial conditions
$$v(q,t_{0})=\tilde{v}(q),\,\,u(q,t_{0})=\tilde{u}(q)$$
has formal solution
\begin{eqnarray*}
v=-\frac{i\hbar}{\mu(t)}\frac{\partial}{\partial q}
\ln\left(\frac{\Psi}{|\Psi|}\right),\,\,u=-\frac{\hbar}{\mu(t)}\frac{\partial}{\partial
q}\ln(|\Psi|),
\end{eqnarray*}
where  $\Psi=\Psi(q,t)$ is a solution of the Schr\"{o}dinger equation
(\ref{evo1}) with general initial condition
$$\Psi(q,t_{0})=\exp(\frac{i\mu(t_{0})}{\hbar}\int^{q}\tilde{v}
(\xi)d\xi)\times\exp(-\frac{\mu(t_{0})}{\hbar}\int^{q}\tilde{u}(\xi)d\xi).$$

\section{Exactly Solvable Nonlinear Models}

 The Caldirola-Kanai model,
\cite{Caldirola},\cite{Kanai}, which is a one dimensional system with an exponentially increasing mass, is the best known model of harmonic oscillator with time-dependent parameters. Here, using the general
discussion in the previous parts, we obtain exact solutions of the
nonlinear problems related with the Caldirola-Kanai oscillator
 \begin{eqnarray}\label{cal-kanai}
\left\{
\begin{array}{ll}
\displaystyle
 i\hbar \frac{\partial \Psi}{\partial
t}=-\frac{\hbar^{2}}{2}e^{-\gamma t}\frac{\partial^{2}
\Psi}{\partial q^{2}}+\frac{1}{2}\omega_{0}^{2}e^{\gamma t}q^{2}\Psi,\\
\displaystyle \Psi(q,0)=\psi(q),\,\,\,\,\,q\in R,
\end{array}
\right.
\end{eqnarray}
 where  $\mu(t)=e^{\gamma t}$ is the integrating
factor, $\Gamma(t)=\gamma$ is the damping term, $\gamma>0$, and
$\omega^{2}(t)=\omega_{0}^{2}$ is a constant frequency.
As known, solutions of the Cadirola-Kanai
oscillator can be found in terms of the solution to the
corresponding classical equation of motion
\begin{equation}\label{clas-cal}
\ddot{x}+\gamma \dot{x}+\omega_{0}^{2}x=0,\,\,\,\,\,x(0)=x_{0}\neq
0,\,\,\,\dot{x}(0)=0.
\end{equation}
Clearly, according to the sign of
 $\Omega^{2}=\omega_{0}^{2}-(\gamma^{2}/4)$ there are three
 different type of behavior- critical damping, under damping and
 over damping. In this article,  we discuss the critical damping case, i.e.  $\Omega^{2}=0.$   If $\Omega^{2}=\omega_{0}^{2}-(\gamma^{2}/4)=0$, then
 the classical equation (\ref{clas-cal})  has solution
\begin{eqnarray*}
x(t)=x_{0}e^{-\frac{\gamma t}{2}}(1+\frac{\gamma}{2}t),
\end{eqnarray*}
and it follows  that
$$g(t)=\frac{-\hbar t}{1+\frac{\gamma}{2}t},\,\,\,\,\,\,\,\,\,\,
R(t)=\left(\frac{e^{\gamma
t}}{(1+\frac{\gamma}{2}t)^{2}+w_{0}^{2}t^{2}}\right)^{1/2}.$$
 Then,
using (\ref{utvar1}),   exact solutions of the
Schr\"{o}dinger equation  (\ref{cal-kanai}) with initial conditions $\Psi(q,0)=\varphi_{k}(q),$ are
\begin{eqnarray*}\label{wave-1}
\Psi_{k}(q,t)&=& N_{k}\left(\frac{e^{\gamma
t}}{(1+\frac{\gamma}{2}t)^{2}+w_{0}^{2}t^{2}}\right)^{1/4}
\exp\left(i(k+\frac{1}{2})\arctan(\frac{-\omega_{0}t}{1+\frac{\gamma}{2}t})\right)\\&
&\times
\exp\left(-i\frac{\omega_{0}^{2}}{2\hbar}\left(\frac{te^{\gamma
t}}{1+\frac{\gamma}{2}t}\right)\left(1-\frac{1}{(1+\frac{\gamma}{2}t)^{2}+w_{0}^{2}t^{2}}\right)q^{2}\right)
\nonumber\\
& &\times \exp\left(-\frac{\omega_{0}}{2\hbar}\left(\frac{e^{\gamma
t}}{(1+\frac{\gamma}{2}t)^{2}+w_{0}^{2}t^{2}}\right)q^{2}\right)\nonumber
H_{k}\left(\sqrt{\frac{\omega_{0}}{\hbar}}\left(\frac{e^{\gamma
t}}{(1+\frac{\gamma}{2}t)^{2}+w_{0}^{2}t^{2}}\right)^{1/2}q\right).
\end{eqnarray*}

\emph{a.  Madelung representation of Caldirola-Kanai Oscillator.}  Madelung representation  of
the wave function decomposes the Schr\"{o}dinger equation
(\ref{cal-kanai}) into a system of  nonlinear coupled  partial
differential equations,
\begin{eqnarray*}\label{hamilton-jacobi-1}
\left\{
\begin{array}{ll}
\displaystyle\frac{\partial S}{\partial t}+\frac{1}{2}e^{-\gamma
t}\left(\frac{\partial S}{\partial
q}\right)^{2}+\frac{\omega_{0}^{2}}{2}e^{\gamma t}q^2=\frac{\hbar^2}{2}e^{-\gamma
t}\left[\frac{1}{\sqrt{\rho}}\frac{\partial^2 \sqrt{\rho}}{\partial
q^2}\right],\\
 \displaystyle
\frac{\partial\rho}{\partial t}+ e^{-\gamma
t}\frac{\partial}{\partial q}\left[\rho\frac{\partial S}{\partial
q}\right]=0.
\end{array}
\right.
\end{eqnarray*}
This system of equations with specific initial conditions
$$S(q,0)=0,\,\,\rho_{k}(q,0)=N_{k}^{2}
\exp\left(-\frac{\omega_{0}}{\hbar}q^{2}\right)H_{k}^{2}\left(\sqrt{\frac{\omega_{0}}{\hbar}}q\right),
$$ has exact solutions of the form
\begin{eqnarray*}
S_{k}(q,t)=\left(-\frac{\omega_{0}^{2}}{2}\left(\frac{te^{\gamma
t}}{1+\frac{\gamma}{2}t}\right)\left(1-\frac{1}{(1+\frac{\gamma}{2}t)^{2}+w_{0}^{2}t^{2}}\right)q^{2}\right)
+\left(\hbar(k+\frac{1}{2})\arctan(\frac{-\omega_{0}t}{1+\frac{\gamma}{2}t})\right),
\end{eqnarray*}
\begin{eqnarray}\label{prob-1}
\rho_{k}(q,t)&=&N_{k}^{2}\left(\frac{e^{\gamma
t}}{(1+\frac{\gamma}{2}t)^{2}+w_{0}^{2}t^{2}}\right)^{1/2}
\exp\left(-\frac{\omega_{0}}{\hbar}\left(\frac{e^{\gamma
t}}{(1+\frac{\gamma}{2}t)^{2}+w_{0}^{2}t^{2}}\right)q^{2}\right)\\
& &\times H_{k}^{2}\left(\sqrt{\frac{\omega_{0}}{\hbar}}\left(\frac{e^{\gamma
t}}{(1+\frac{\gamma}{2}t)^{2}+w_{0}^{2}t^{2}}\right)^{1/2}q\right)\nonumber.
\end{eqnarray}

\emph{b. Hydrodynamic equations}. The  system of  hydrodynamic equations  for  the velocity and density
of the fluid
\begin{eqnarray*}\label{clas-velocity-1}
\left\{ \begin{array}{ll}
\displaystyle
 \frac{\partial v}{\partial
t}+\gamma v + v \frac{\partial v}{\partial q}=-e^{-\gamma
t}\frac{\partial}{\partial
q}\left[\frac{-\hbar^{2}e^{-\gamma
t}}{2\sqrt{\rho}}\frac{\partial^2 \sqrt{\rho}}{\partial
q^{2}}+\frac{\omega_{0}^{2}}{2}e^{\gamma
t}q^2 \right]\\
\displaystyle \frac{\partial\rho}{\partial
t}+\frac{\partial}{\partial q}\left[\rho v\right]=0,
\end{array}
\right.
\end{eqnarray*}
with specific initial conditions
\begin{eqnarray*}\label{ic2-1}
v(q,0)=0,\,\,\rho_{k}(q,0)=N_{k}^{2}
\exp\left(-\frac{\omega_{0}}{\hbar}q^{2}\right)H_{k}^{2}\left(\sqrt{\frac{\omega_{0}}{\hbar}}q\right).
\end{eqnarray*}
 has solutions
\begin{eqnarray*}\label{vk-1}
v(q,t)=\left(-\omega_{0}^{2}\left(\frac{t}{1+\frac{\gamma}{2}t}\right)\left(1-\frac{1}{(1+\frac{\gamma}{2}t)^{2}+w_{0}^{2}t^{2}}\right)q\right)
\end{eqnarray*}
and $\rho_{k}(q,t)$ given by (\ref{prob-1}).

\vspace{.2in}

 \emph{c. The potential Schr\"{o}dinger-Burgers equation. } The IVP for potential Schr\"{o}dinger-Burgers equation
\begin{eqnarray}\label{pot-burgers-1}
\left\{
\begin{array}{ll}
\displaystyle\frac{\partial F}{\partial
t}+\gamma F+\frac{1}{2}(\frac{\partial F}{\partial q})^{2}+\frac{\omega_{0}^{2}}{2}q^{2}=\frac{i\hbar}{2}e^{-\gamma t}(\frac{\partial^{2}F}{\partial
q^{2}}),\\
F_{k}(q,0)=i
\left(\frac{\omega_{0}}{2}q^{2}-\hbar\ln(N_{k}H_{k}(\sqrt{\frac{\omega_{0}}{\hbar}}q))\right),\,\,\,\,\,\,k=0,1,2,...
\end{array}
\right.
\end{eqnarray}
has  solutions  $F_{k}(q,t)=\textsf{F}_{1,k}(q,t)+i\textsf{F}_{2,k}(q,t),$
where
\begin{eqnarray*}
\textsf{F}_{1,k}=-\frac{\omega_{0}^{2}}{2}\left(\frac{t
}{1+\frac{\gamma}{2}t}\right)\left(1-\frac{1}{(1+\frac{\gamma}{2}t)^{2}+w_{0}^{2}t^{2}}\right)q^{2}
+\hbar e^{-\gamma
t}(k+\frac{1}{2})\arctan(\frac{-\omega_{0}t}{1+\frac{\gamma}{2}}),
\end{eqnarray*}
\begin{eqnarray*}
\textsf{F}_{2,k}
&=&\frac{\omega_{0}}{2}\left(\frac{1}{(1+\frac{\gamma}{2}t)^{2}+w_{0}^{2}t^{2}}\right)q^{2}
-\hbar e^{-\gamma t}\ln\left[ N_{k}\left(\frac{e^{-(\gamma/2)
t}}{(1+\frac{\gamma}{2}t)^{2}+w_{0}^{2}t^{2}}\right)^{1/2}\right]\\
& &-\hbar e^{-\gamma t}\ln
H_{k}\left(\sqrt{\frac{\omega_{0}}{\hbar}\left(\frac{e^{\gamma
t}}{(1+\frac{\gamma}{2}t)^{2}+w_{0}^{2}t^{2}}\right)}\,\,\,q\right).
\end{eqnarray*}

\emph{d. The Schr\"{o}dinger-Burgers equation.} The IVP for the non-linear complex Schr\"{o}dinger-Burgers equation
\begin{eqnarray}\label{burgers-cal-1}
\left\{
\begin{array}{ll}
\displaystyle \frac{\partial V}{\partial t}+\gamma V+V\frac{\partial
V}{\partial q}+\omega_{0}^{2}q =i\hbar\frac{ e^{-\gamma
t}}{2}\frac{\partial^{2}V}{\partial q^{2}},\\
\displaystyle V_{k}(q,0)=i\left[\omega_{0}q-2k\sqrt{\hbar\omega_{0}}\left(
\frac{H_{k-1}(\sqrt{\frac{\omega_{0}}{\hbar}}q)}{H_{k}(\sqrt{\frac{\omega_{0}}{\hbar}}q)}\right)\right],
\end{array}
\right.
\end{eqnarray}
has solutions
\begin{eqnarray}\label{vqt}
V_{k}(q,t) =
\left(-\omega_{0}^{2}\left(\frac{t}{1+\frac{\gamma}{2}t}\right)\left(1-\frac{1}
{(1+\frac{\gamma}{2}t)^{2}+w_{0}^{2}t^{2}}\right)q\right)
+i\left(\frac{\omega_{0}}{(1+\frac{\gamma}{2}t)^{2}+w_{0}^{2}t^{2}}\right)q \nonumber\\
-i2k \sqrt{\hbar\omega_{0}}e^{-\gamma t} \left(\frac{e^{\gamma
t}}{(1+\frac{\gamma}{2}t)^{2}+w_{0}^{2}t^{2}}\right)^{1/2}\times
\left(\frac{H_{k-1}(\sqrt{\frac{\omega_{0}}{\hbar}\left(\frac{e^{\gamma
t}}{(1+\frac{\gamma}{2}t)^{2}+w_{0}^{2}t^{2}}\right)}\,\,\,q)}{H_{k}(\sqrt{\frac{\omega_{0}}{\hbar}\left(\frac{e^{\gamma
t}}{(1+\frac{\gamma}{2}t)^{2}+w_{0}^{2}t^{2}}\right)}\,\,\,q)}\right).
\end{eqnarray}

 The complex velocity function, written in the form $
 V(q,t)=v(q,t)+iu(q,t),$ where $v,u$ are real-valued functions,  splits the
 Schr\"{o}dinger-Burgers
equation (\ref{burgers-cal-1}) into the system
\begin{eqnarray*}\label{hyd-eq-cal}
\left\{
\begin{array}{ll}
\displaystyle \frac{\partial v}{\partial t}+\gamma v+v\frac{\partial
v}{\partial q}-u\frac{\partial u}{\partial q}+\omega^{2}_{0}q=-
\frac{\hbar}{2}e^{-\gamma t}\frac{\partial^{2} u}{\partial
q^{2}}\,\,,\\
\displaystyle \frac{\partial u}{\partial t}+\gamma u+u\frac{\partial
v}{\partial q}+u\frac{\partial v}{\partial
q}=\frac{\hbar}{2}e^{-\gamma t}\frac{\partial^{2}v}{\partial q^{2}}.
\end{array}
\right.
\end{eqnarray*}
This system with specific initial conditions
$$v(q,0)=0,\,\,
u_{k}(q,0)=\omega_{0}q-2k\sqrt{\hbar\omega_{0}}\left(\frac{H_{k-1}
(\sqrt{\frac{\omega_{0}}{\hbar}}q)}{H_{k}(\sqrt{\frac{\omega_{0}}{\hbar}}q)}\right),$$
clearly, has solutions $v(q,t)$ and $u_{k}(q,t)$ which are respectively, the real and imaginary parts of $V_{k}(q,t)$ in expression (\ref{vqt}).

\emph{e. Motion of Zeros and Poles,  $\Omega^{2}=0$.} From expression (\ref{wave-1}) we see that the solution
$\Psi_{k}(q,t)$ (also $\rho_{k}(q,t)$ ) of the linear
Schr\"{o}dinger equation (\ref{cal-kanai}) has zeros at points where
$$ H_{k}(\sqrt{\frac{\omega_{0}}{\hbar}\left(\frac{e^{\gamma
t}}{(1+\frac{\gamma}{2}t)^{2}+w_{0}^{2}t^{2}}\right)}\,\,\,q)=0,$$
and these zeros are pole singularities of the solution $V_{k}(q,t)$ (also
$|V_{k}(q,t)|^{2}$ ) for the nonlinear Schr\"{o}dinger-Burgers
equation (\ref{burgers-cal-1}). Therefore,
for each $k=0,1,2,3,...,$ the motion of the zeros and poles is
described by the curves
\begin{equation}\label{zeros}
q_{k}^{(l)}(t)=\tau_{k}^{(l)}\sqrt{\frac{\hbar}{\omega_{0}}}\times
e^{-\frac{\gamma}{2}t}\times
\sqrt{(1+\frac{\gamma}{2}t)^{2}+w_{0}^{2}t^{2}},
\end{equation}
where $\tau_{k}^{(l)},\,\,$$l=1,2,...k,\,\,$  are the
zeros of the Hermite polynomial $H_{k}(\xi).$
Clearly, at initial time the position of the zeros and poles is
$q_{k}^{(l)}(0)=\tau_{k}^{(l)}\sqrt{\hbar/\omega_{0}},\,$ and
when $\gamma>0,$ $\,t\rightarrow\infty$ one has
$q_{k}^{(l)}(t)\rightarrow 0$ due to increasing mass
$\mu(t)=e^{\gamma t},$ (dissipation).

 In  the figure we illustrate  the probability density
 function $\rho_{2}(q,t)$ for Caldirola-Kanai oscillator, which shows Dirac-delta behavior at
 time infinity and the position of the
  moving zeros and poles $q_{2}^{(1)}(t)=-\frac{e^{-t}}{\sqrt{2}}\sqrt{(1+t)^{2}+t^{2}}$,
$q_{2}^{(2)}(t)=-q_{2}^{(1)}(t).$  For the plots, constants are
chosen as $x_{0}=\hbar=\omega_{0}=1$ and $\gamma=2.$

\begin{figure}[h]
\begin{center}
\epsfig{figure=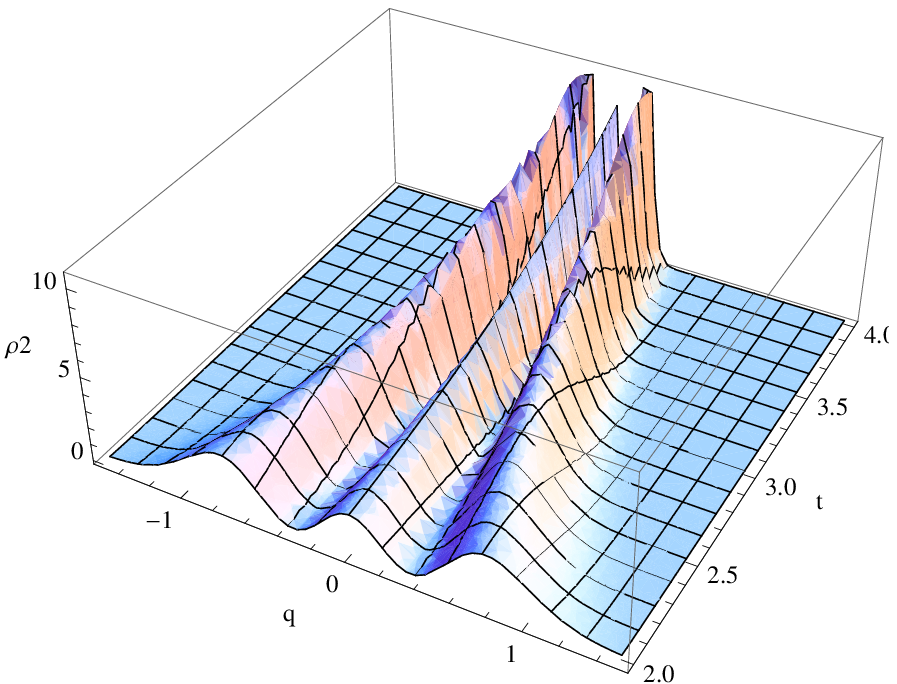,height=5.5cm,width=7cm}
\hspace{.2cm}
\epsfig{figure=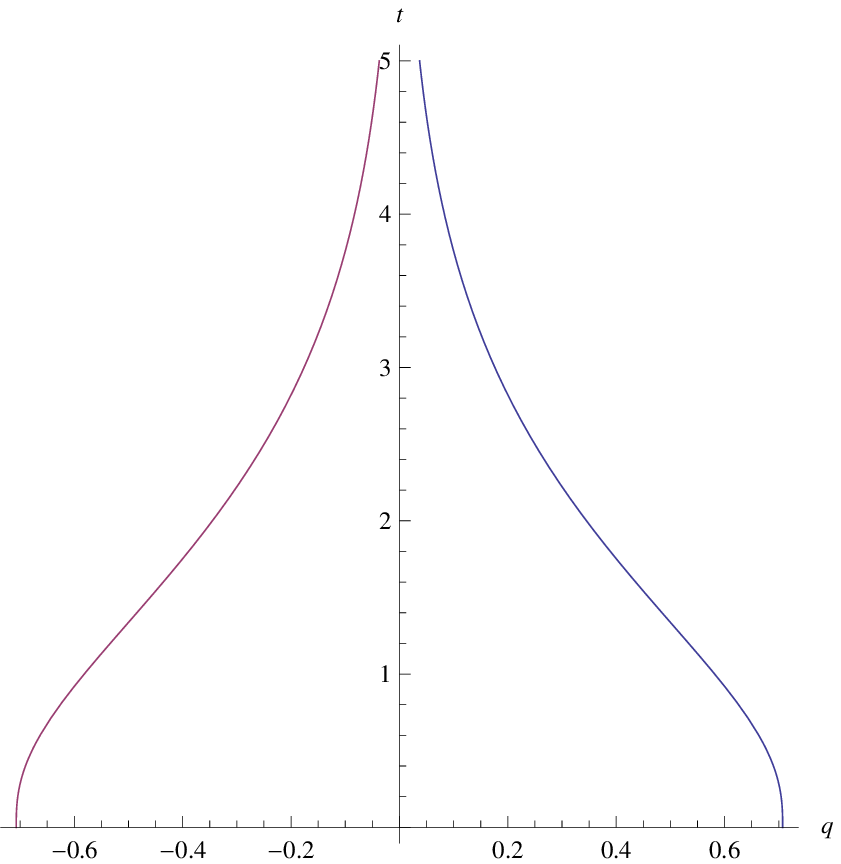,height=4.5cm,width=4.5cm}
\end{center}
\caption{Case $\Omega^{2}=0.$ a) Evolution of
probability density $\rho_{2}(q,t).$ b) Plot of moving zeros and
poles.} 
\end{figure}

\vspace{3in}

\section{Conclusion} In the present paper we have described time variable Madelung fluid and its linearization
in terms of time variable linear Schr\"odinger equation. Our model,
as  descriptive  of dissipative quantum fluid, admits exact solution
for  specific time dependent systems, like the harmonic oscillator
with time dependent frequency and mass. In this case, exact time
evolution has been described in terms of solutions for the
corresponding damped classical oscillator. In particular, for the
damping simulated by an exponentially growing mass (the
Caldirola-Kanai model), it can be shown that the quantum damping
squeezes the density function of the fluid and leads to Dirac-delta
function. This can have some implications in quantum cosmology. In
fact, if $\Psi(q,t)$ is a solution of  the Caldirola-Kanai
oscillator for a damped system
\begin{eqnarray*}
i\hbar \frac{\partial \Psi}{\partial
t}=-\frac{\hbar^{2}}{2}e^{-\gamma t}\frac{\partial^{2}
\Psi}{\partial q^{2}}+\frac{1}{2}\omega_{0}^{2}e^{\gamma
t}q^{2}\Psi,
\end{eqnarray*}
then $\,\,\Phi(q,t)=\Psi^{*}(q,-t),\,\,$  where $(*)$ denotes
complex conjugation, is a solution of the related dual amplified
system
\begin{eqnarray*}
i\hbar \frac{\partial \Phi}{\partial
t}=-\frac{\hbar^{2}}{2}e^{\gamma t}\frac{\partial^{2} \Phi}{\partial
q^{2}}+\frac{1}{2}\omega_{0}^{2}e^{-\gamma t}q^{2}\Phi.
\end{eqnarray*}
 Hence, knowing that the solution $\Psi(q,t)$ of the Caldirola-Kanai model  has merging zeros and describes collapse of the wave function to Dirac-delta function at time infinity, leads to possible interpretation of the solution $ \Psi^{*}(q,-t)$ for the dual system. Namely, we will have expanding wave function with creation of zeros as point particles from initial singularity at time zero. And this evolution simulates quantum mechanism similar to creation of expanding Universe from initial singularity in Big-Bang cosmology.

Finally, we note that by our results it is possible to find explicit
exact solutions to a wide class of exactly solvable dissipative
quantum fluids and complex Schr\"{o}dinger-Burgers equations. For
this we can use our recent work on exactly solvable dissipative
systems, such as quantum Sturm-Liouville problems  Ref.\cite{S.O.E}.
Then, it is possible also to describe the dynamics of the zeros and
poles in the corresponding dissipative linear and nonlinear systems.
These questions are under investigation now.

\end{document}